\documentclass[twocolumn,showpacs,preprintnumbers,amsmath,amssymb,aps,epsfig,psfig,superscriptaddress,showkeys,email]{revtex4}
\usepackage{setspace}
\usepackage{color} 
\usepackage{graphics}
\usepackage{graphicx}
\usepackage{supertabular}
\usepackage{longtable}
\begin{document}
\title{\bf On the relation between native geometry and conformational plasticity}
\author{Patr\'icia  F.~N. Fa\'\i sca}
\affiliation{Centro de F\'\i sica Te\'orica e Computacional, Universidade 
de Lisboa, Av. Prof. Gama Pinto 2, 1649-003 Lisboa, Portugal}
\email{patnev@cii.fc.ul.pt}
\author{Cl\'audio M. Gomes}
\affiliation{Instituto de Tecnologia Qu\' imica e Biol\'ogica,
Universidade Nova de Lisboa, Av. da Republica EAN, 2785-572 Oeiras, Portugal}

\pacs{{\bf 87.15.Cc; 87.15.ak; 87.15.hm; 87.15.hp}}
\keywords{\bf{mutagenesis, $\phi$-value analysis, folding pathways, long-range contacts}}

\begin{abstract}
In protein folding the term plasticity refers to the number of alternative folding pathways 
encountered in response to free energy perturbations such as those induced by mutation. Here 
we explore the relation between folding plasticity and a gross, generic feature of the native geometry, namely, the relative number of local and non-local native contacts. The results from our study, which is based on Monte Carlo simulations of simple lattice proteins, show that folding to a structure that is rich in local contacts is considerably more plastic than folding to a native geometry characterized by having a very large number of long-range contacts (i.e., contacts between amino acids that are separated by more than 12 units of backbone distance). The smaller folding plasticity of `non-local' native geometries is probably a direct consequence of their higher folding cooperativity that renders the folding reaction more robust against single- and multiple-point mutations. 
\end{abstract}
\maketitle

\section{Introduction}
During the last 15 years significant progress has been achieved towards a complete understanding of the kinetics and mechanisms of protein folding. The synergistic link between computer simulations and {\it in vitro} experiments has proven particularly fruitful in this endeavour~\cite{DOBSON_KARPLUS, SHEA}. 
Much of our current knowledge on protein folding has been gathered by studying small (i.e., with less than 100 amino acids) monomeric proteins, epitomized by the likes of the 64-residue protein Chymotrypsin Inhibitor 2 (CI2). Indeed, their two-state folding kinetics~\cite{JACKSON} renders them particularly suitable models to investigate, both {\it in vitro} and {\it in silico}, the rather complex phenomenon that is the folding `reaction'. \par   
A major and challenging task in studying  two-state proteins is the structural characterization of the folding transition state (TS), located on the top of the free energy barrier that separates the native fold from the ensemble of unfolded conformations. Indeed, due to its fleeting nature,  the commonly available biophysical tools have revealed inappropriate to probe the TS' structure. Thus, experimental studies of the TS have remained predominantly rooted in the use of a particular class of protein engineering methods, the so-called $\phi$-value analysis, developed by Fersht and coworkers back in the late 1980s~\cite{FERSHT_BOOK}. In the $\phi$-value analysis a non-disruptive mutation (i.e., a mutation that does not change the structure of the native state, and does not alter the folding pathway either) is made at some position in the protein sequence~\cite{FERSHT}. The change in the activation energy of folding of the mutant with respect to that of the wild-type (WT) protein, denoted by $\Delta \Delta G^{TS-D}$, is measured together with the change in the free energy of folding,  $\Delta \Delta G^{N-D}$, caused by the mutation.  The corresponding $\phi$ value is then defined as the ratio between these quantities, namely, $\phi = \Delta \Delta G ^{TS-D}/ \Delta\Delta G ^{N-D}$. A $\phi$ value of unity means that the energy of the TS is perturbed upon mutation exactly as much as that of the native state, which has ''traditionally´´ been interpreted as if the protein structure is folded at the site of mutation in the TS. Conversely, residues that are unfolded in the TS exhibit $\phi$-values of zero. 

The traditional interpretation of fractional $\phi$ values is, however, not straightforward as they might indicate the existence of multiple folding pathways or a unique TS with genuinely weakened interactions~\cite{FERSHT_BOOK}. Moreover, the so-called nonclassical $\phi$-values ($\phi >1$ or $\phi < 0$ ) are difficult to interpret in the traditional $\phi$-value model. Recently, Weikl and co-workers have proposed a new model that is able to capture and interpret nonclassical $\phi$-values~\cite{WEIKL_PNAS, WEIKL_JMB}. The model assumes that cooperative structural elements (e.g., and $\alpha$-helix or a $\beta$-hairpin) are fully formed in the TS, and that a mutation on a single residue affects the whole structural element where the native contacts established by the mutated residue are located. Likewise, the activation energy of folding in the definition of the $\phi$-values contain explicit free energy contributions from different substructural elements of the protein. This model was recently applied to study two small $\beta$-sheet domains, namely the FBP and PIN WW domains, which were assumed to fold via two distinct folding pathways (and transition states), and it was able to reproduce $\phi$-values in good agreement with those reported experimentally~\cite{WEIKL_BJ}.\par 
The concept of conformational/folding plasticity~\cite{MUNOZ}, or folding malleability~\cite{OLIVEBERG_REV}, refers to the number of alternative folding pathways encountered in response to free energy perturbations such as those induced by solvent changes or mutation; the larger that number, the broader the TS, and the larger the plasticity. 
Although it is experimentally very difficult to investigate in a direct manner the number of folding pathways leading to the native state, evidence for the existence of conformational plasticity was found in a few cases by means of protein engineering experiments. For example, single point mutations in the N-terminal domain of the monomeric $\lambda_{6-85}$-repressor revealed a broad TS~\cite{OAS}. The use of circular permutation methods - a less milder protein engineering technique where the protein's backbone is cut-up at some point, and its original N- and C-terminal parts are linked together in a process that leads to minimal alteration of the native fold - revealed considerable structural changes in the TSs of both the $\alpha$-spectrin SH3 domain~\cite{SERRANO} and src SH3 domain~\cite{BAKER_1}, and even more dramatic modifications in the TS of the ribosomal protein S6~\cite{OLIVEBERG}, a finding that was corroborated by computer simulation studies~\cite{HUBNER, HUBNER_1, CLEMENTI}. The plasticity exhibited by these proteins has been taken has evidence for the existence of multiple folding pathways~\cite{MUNOZ}.\par 
In a recent study, Klimov and Thirumalai have studied the influence of the distribution of secondary structural elements along the protein sequence using an off-lattice coarse grained model and Langevin Molecular Dynamics simulations. The major result that came out of their study is that a symmetric distribution of $\alpha$ helices and $\beta$ sheets with respect to sequence midpoint favours multiple folding pathways~\cite{KLIMOV_JMB}.\par 
A study of the validity of the $\phi$-value analysis as a tool for identifying critical (i.e., nucleating) residues in protein folding and providing a structural characterization of the TS was recently presented in~\cite{FAISCA_2008}. This study involved measuring the folding time (i.e., the inverse of the folding rate) of all possible single point mutations and many double point mutations in two model proteins with different native geometries. Here we propose to explore the link between conformational plasticity and a gross geometric trait of the native structure, the relative number of local and non-local native contacts, by using as starting point the mutational data reported in~\cite{FAISCA_2008}. For each native geometry, we selected a number of mutations with different kinetic effects. We investigate the relation between folding plasticity and native geometry by studying the impact of these mutations on the folding `reaction' from the point of
view of the conformational changes it encompasses. This is done by monitoring for each mutant the degree of nativeness of each residue along the folding process and comparing it with the wild type protein.
Our analysis shows that folding to the native geometry that is rich in nonlocal long-range (LR) contacts is considerably less plastic (i.e.,  more conformationally robust) than folding to the model protein that has predominantly local contacts, and that this is possibly a direct consequence of the more cooperative folding transition exhibited by the non-local geometry.  
This reinforces the results of~\cite{FAISCA_2008}, according to which  
the picture of the TS emerging from the $\phi$-value analysis is  more reliable when applied to target proteins having a distinctively large number of non-local native contacts.
Therefore, if $\phi$-value mutational data is to be interpreted in the traditional way, which is so far the most commonly adopted view, it is important to select native folds where the number of LR native contacts is sharply dominant; they show a higher robustness against mutation and, at the coarse level of contact cluster, they tend to fold in a Levinthal-like manner, i.e., as single route folding proteins.\par
This article is organized in the following manner. After the introductory section, the protein models and simulation methodologies are described. Then, we present and discuss the results from simulations. Finally, in the last section we draw some concluding remarks.
         
\begin{figure*}
{\rotatebox{270}{\resizebox{6cm}{5cm}{\includegraphics{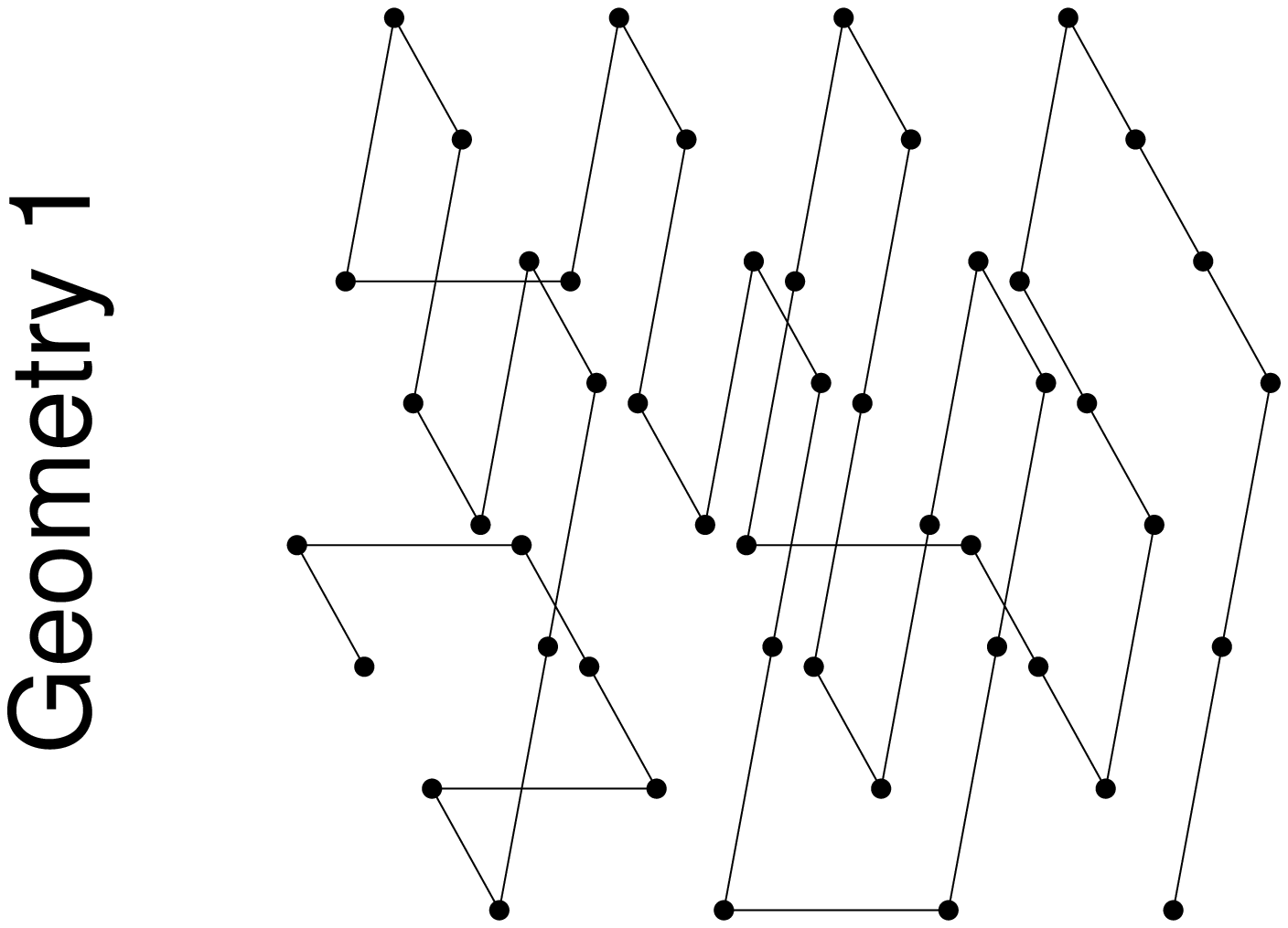}}}} 
\hspace{1cm}
{\rotatebox{270}{\resizebox{6cm}{5cm}{\includegraphics{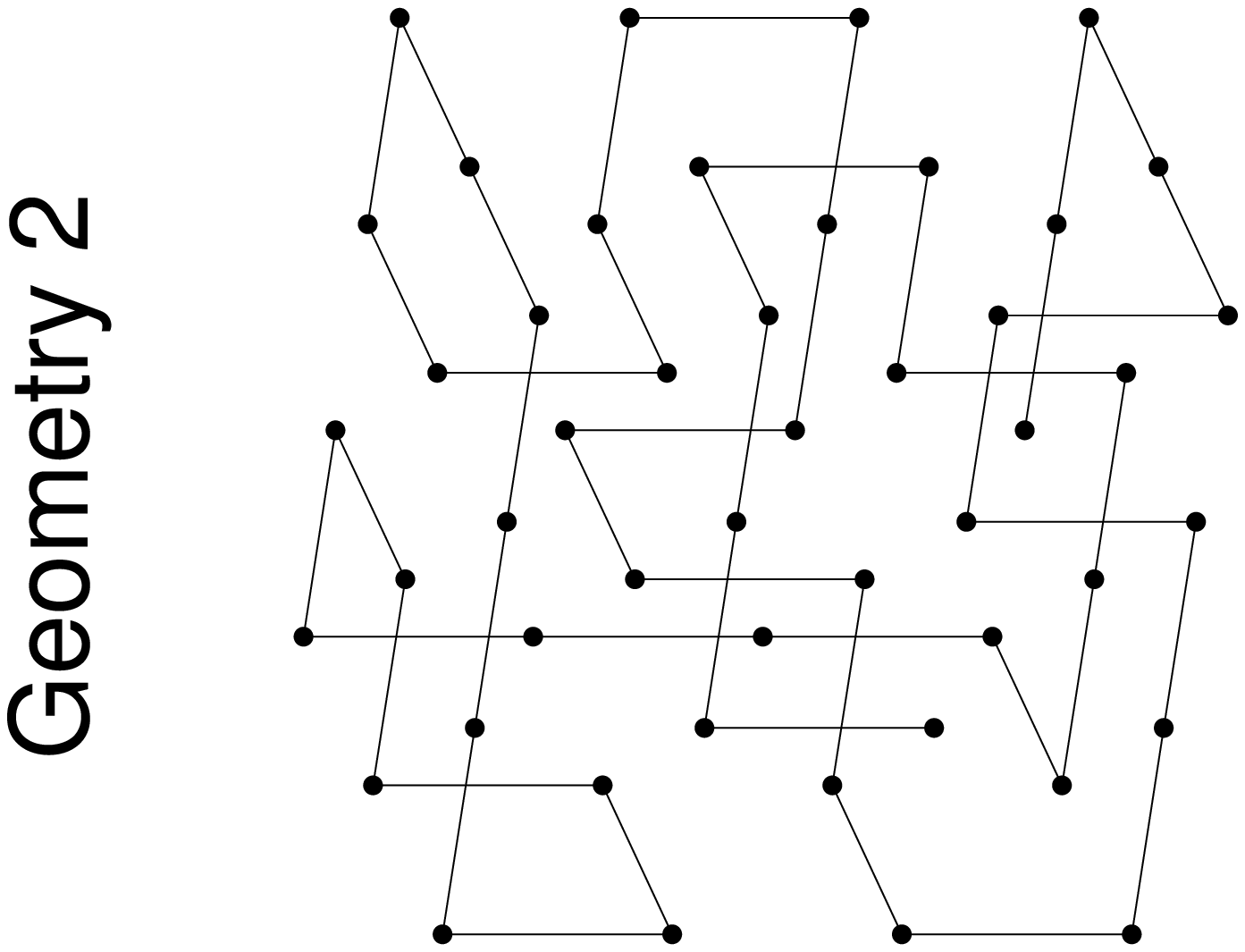}}}} \\
{\rotatebox{0}{\resizebox{7cm}{7cm}{\includegraphics{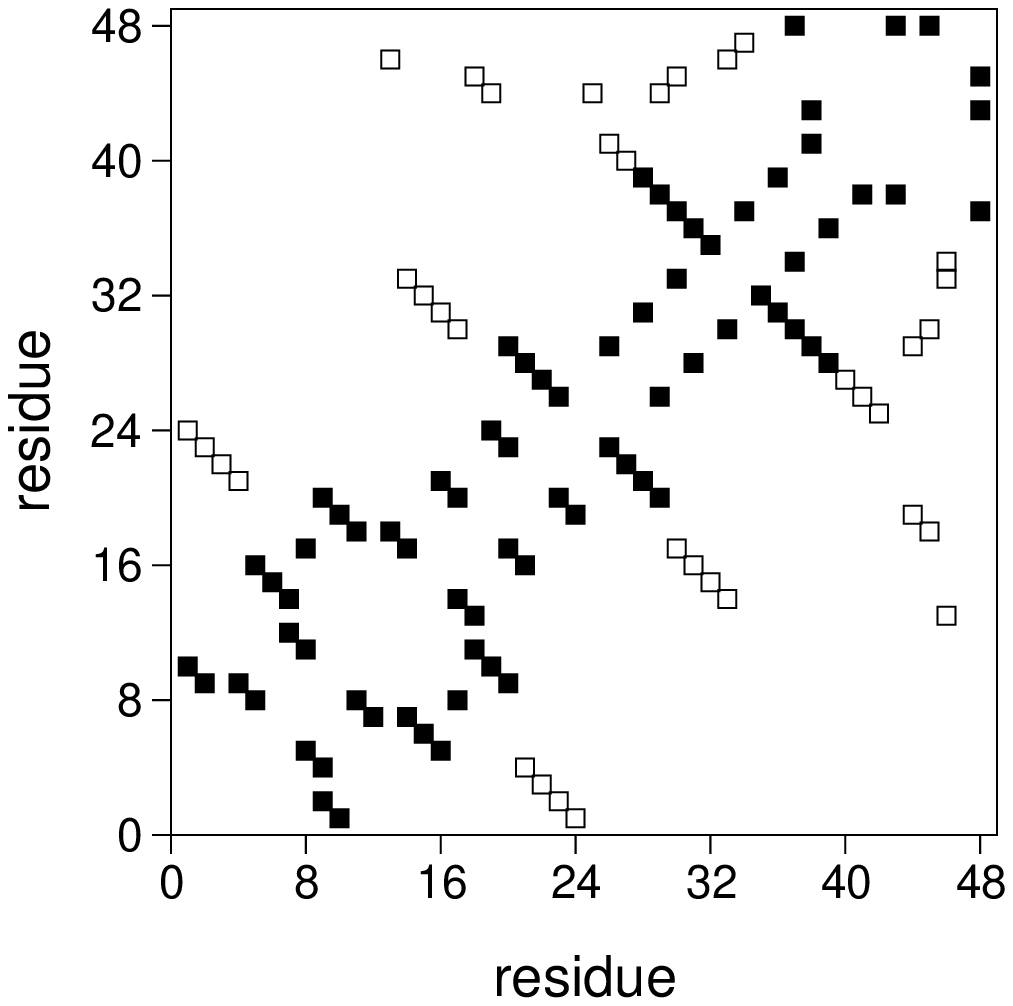}}}}
{\rotatebox{0}{\resizebox{7cm}{7cm}{\includegraphics{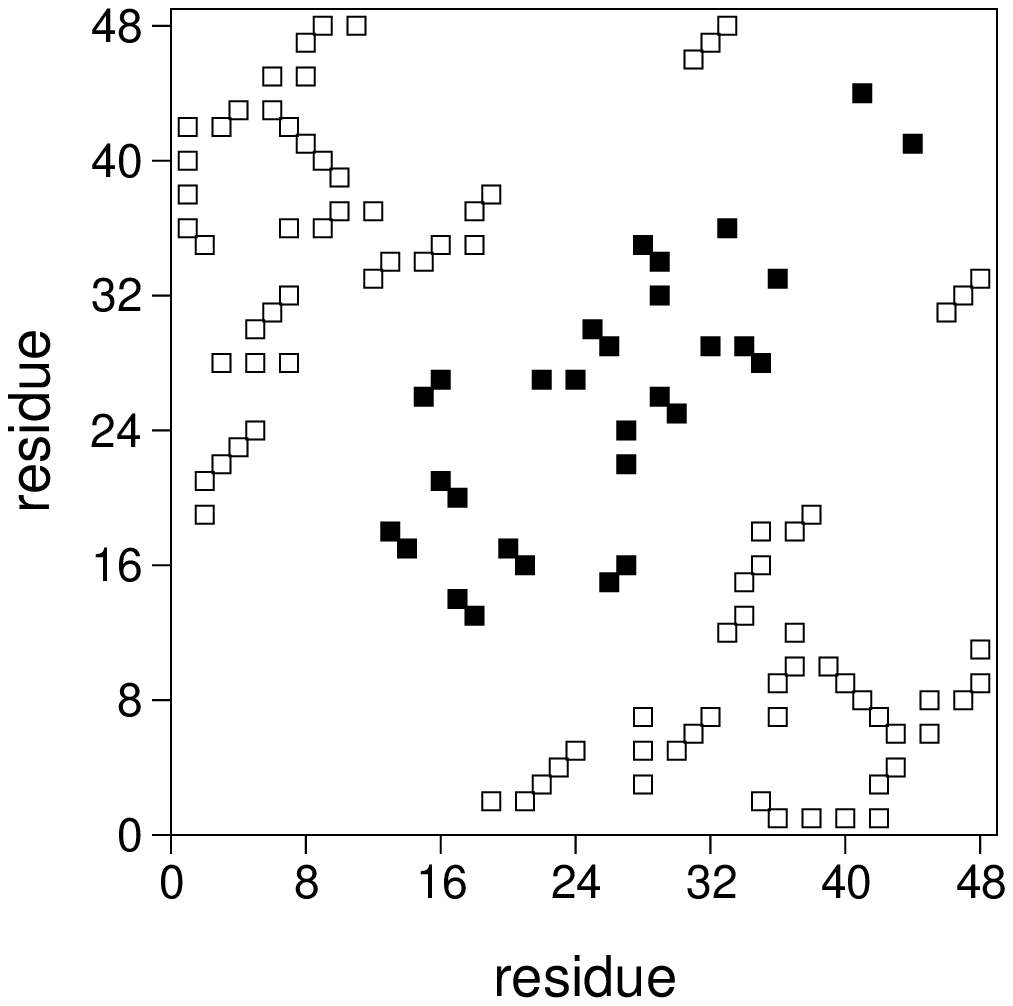}}}}
\caption{{\bf Three dimensional representation of geometry 1 and 2 (top row) and corresponding contact maps (bottom row)}. Each square in the contact map represents a native contact. For structures that like ours are maximally compact cuboids with $N=48$ residues there are 57 native contacts. A non-local contact between two residues $i$ and $j$ is defined as LR if their sequence separation is at least 12 units, i.e., $|i-j| \geq 12$~\cite{GROMIHA}. Accordingly, the number of LR (white squares) contacts in geometry 1 is 19 and in geometry 2 is 42.}
\label{fig:no1}
\end{figure*}

\section{Models and Methods}

\subsection{The G\={o} model and simulation details}

We consider a simple three-dimensional lattice model of a protein molecule with
chain length $N$=48. In such a minimalist model amino acids, represented by beads of uniform size, occupy the lattice vertices and the peptide bond, which covalently connects amino acids along the polypeptide chain, is represented by sticks with uniform (unit) length corresponding to the lattice spacing. \par
To mimic protein energetics we use the G\={o} model~\cite{GO}. In the G\={o} model the energy of a conformation, defined by the set of bead coordinates $\lbrace \vec{r_{i}} \rbrace$, is given by the contact Hamiltonian  
\begin{equation}
H(\lbrace \vec{r_{i}} \rbrace)=\sum_{i>j}^N
\epsilon \Delta(\vec{r_{i}}-\vec{r_{j}}),
\label{eq:no1}
\end{equation}
where the contact function $\Delta (\vec{r_{i}}-\vec{r_{j}})$, is unity only 
if beads $i$ and $j$ form a non-covalent native contact, i.e., a contact 
between a pair of beads that is present in the native structure, and is zero otherwise. 
The G\={o} potential is based on the idea that the native fold is very 
well optimized energetically. Accordingly, it ascribes equal stabilizing 
energies (e.g., $\epsilon=-1.0 $) to all the native contacts and neutral energies 
($\epsilon =0$) to all non-native contacts. 
The motivation to use the G\={o} model is based on the well accepted finding that for small, single domain
two-state proteins, the geometry of the native fold is the major determinant of folding kinetics~\cite{PLAXCO, GROMIHA, ZHOU}.\par
In order to mimic the protein's relaxation towards the native state we use a standard 
Monte Carlo (MC) algorithm together with the kink-jump move set~\cite{BINDER}. 
Accordingly, local random displacements of one or two beads (at the same time) are repeatedly accepted or rejected in accordance with the standard Metropolis MC rule~\cite{METROPOLIS}.
A MC simulation starts from a randomly generated unfolded conformation 
and the folding dynamics is monitored by following the evolution of the fraction of native
contacts, $Q=q/L$, where $L$ is number of contacts in the native fold and $q$ is the 
number of native contacts formed at each MC step. The number of MC steps required to fold to the native state (i.e., to achieve $Q=1.0$) is the first passage time (FPT) and the folding time, $t$, is computed as the mean FPT of 100 simulations. Except otherwise stated folding is studied at the so-called optimal folding temperature, the temperature that minimizes the folding time~\cite{OLIVEBERG2, JCPSHAKH, CIEPLAK, PFN1}. The folding transition temperature, T$_{f}$, is defined is the temperature at which denatured states and the native state are equally populated at equilibrium. In the context of a lattice model it can be defined as the temperature at which the average value $<Q>$ of the fraction of native contacts is equal to 0.5~\cite{ABKEVICH1}. In order to determine T$_f$ we averaged $Q$, after collapse to the native state, over MC simulations lasting $\sim$ 10$^9$ MCS. \par

\subsection{Target geometries}

Two native folds, which are amongst the `simplest' (geometry 1) and the most `complex' (geometry 2) cuboid geometries found through lattice simulations of homopolymer relaxation, were considered in this study (Figure~\ref{fig:no1}). A non-local contact is considered long-range (LR) if the two beads participating in the contact are separated by more than 12 units of backbone distance~\cite{GROMIHA}. Accordingly, 33\% of the native contacts in geometry 1 are LR, while geometry 2 has 74\% LR contacts.

\section{Results}

\subsection{A picture of the folding reaction}

The mechanistic equivalent of the experimental $\phi$-value of residue $i$ at time $t$, $\phi_{i}^{mec}(t)$ , is given by the number of native contacts $q_{i}^{\Gamma} (t)$, the residue establishes in conformation $\Gamma$, normalised to the number of contacts it establishes in the native fold, $q^{native}$, i.e., $\phi_{i}^{mec}= q_{i}^{\Gamma}(t)/q^{native}$~\cite{VENDRUSCOLO, HUBNER_JMB, PACI_07}; a residue is fully native if $\phi^{mec}=1$. Thus, $\phi^{mec}$ provides a {\it direct} measure of structure formation. \par
In what follows we use $\phi^{mec}$ to obtain a picture of the folding reactions leading to geometry 1 and geometry 2, and we use the fraction of native contacts as a progress variable. In particular, for fraction of native contacts, $Q$, we compute the number of times each residue is fully native (i.e., $\phi^{mec}=1$) and normalize it to the total number of times $Q$ is counted in the course of a folding event. The probability thus computed is then averaged over an ensemble of 100 independent folding simulations. We remark that although the probability to fold, $P_{fold}$, is the most accurate progress variable in lattice simulations of protein folding~\cite{DU}, its use for the purposes of the present goal is computationally prohibitive as it would have to be evaluated for {\it every} conformation sampled in a folding event. This is why we use instead the fraction of native contacts, which was recently shown to measure correctly the degree of closeness to the native fold~\cite{CHO} for proteins with smooth energy landscapes (i.e., single exponential kinetics)~\cite{GILLESPIE, CLEMENTI1}, like those considered here.\par
\begin{figure*}
{\rotatebox{0}{\resizebox{5cm}{5.0cm}{\includegraphics
{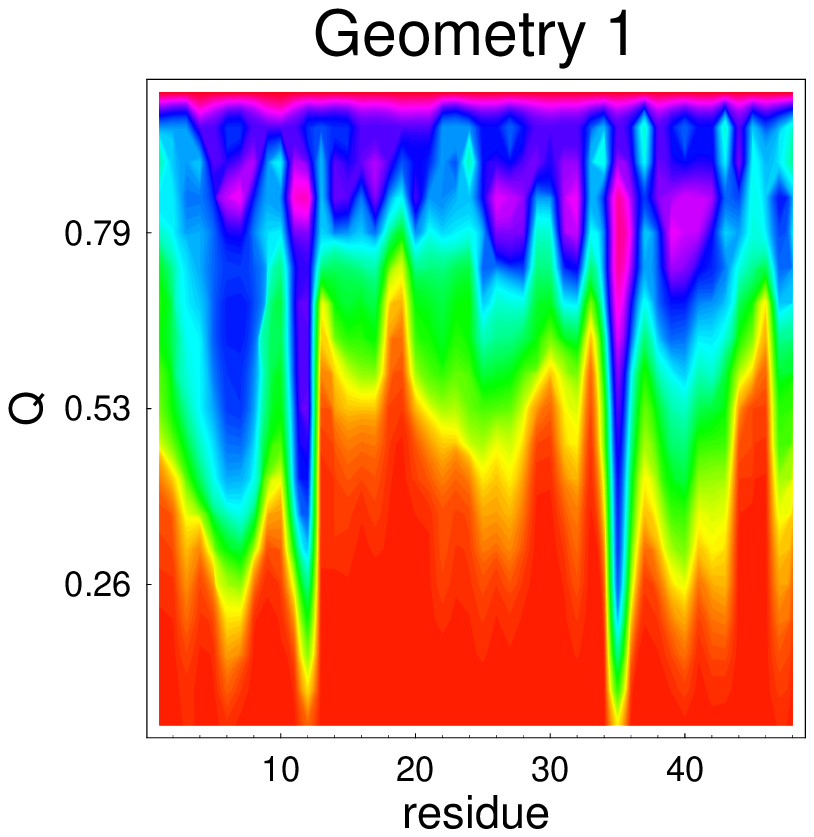}}}\hspace{0.7cm}} 
{\rotatebox{0}{\resizebox{5cm}{5.0cm}{\includegraphics
{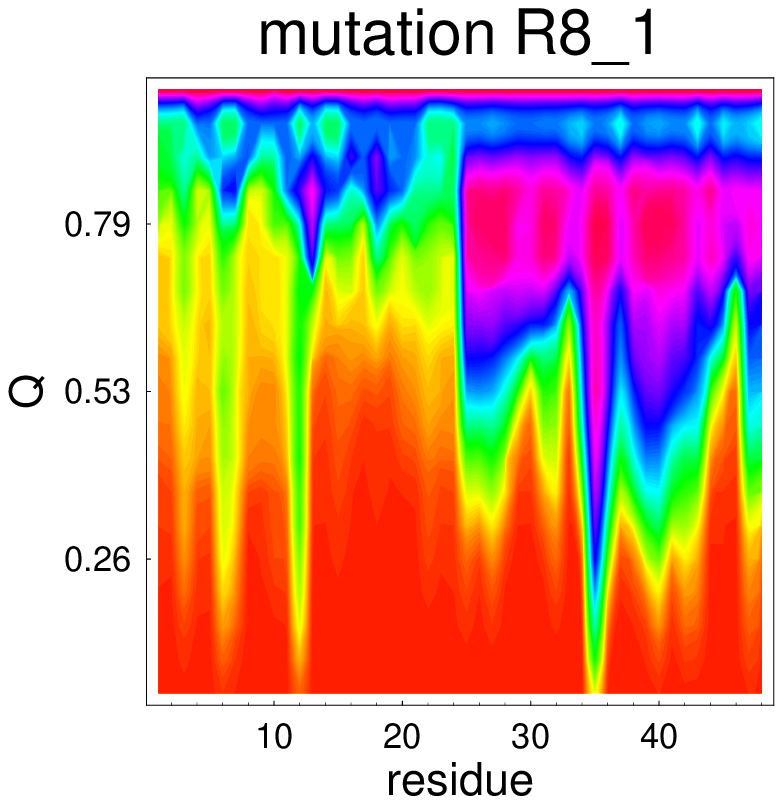}}}} \\
{\rotatebox{0}{\resizebox{5cm}{5.0cm}{\includegraphics
{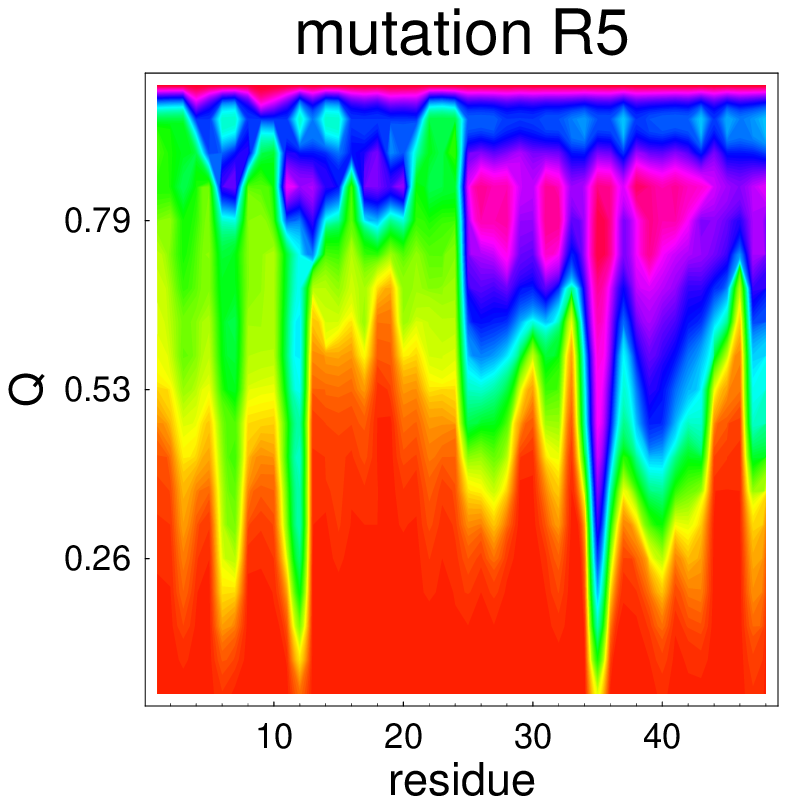}}}} \hspace{0.7cm}
{\rotatebox{0}{\resizebox{5cm}{5.0cm}{\includegraphics
{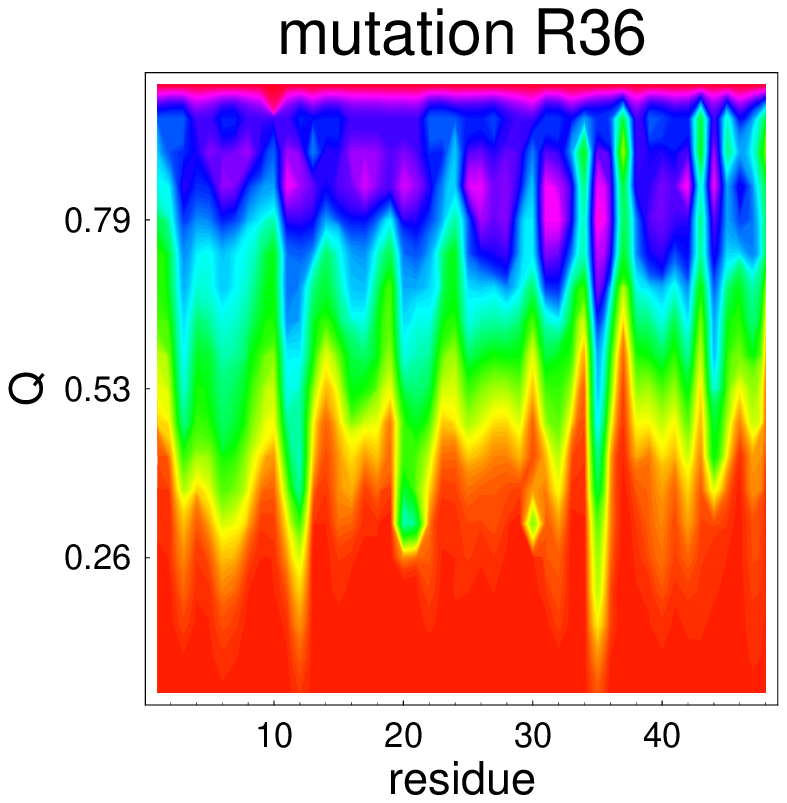}}}} \\
{\rotatebox{0}{\resizebox{5cm}{5.0cm}{\includegraphics
{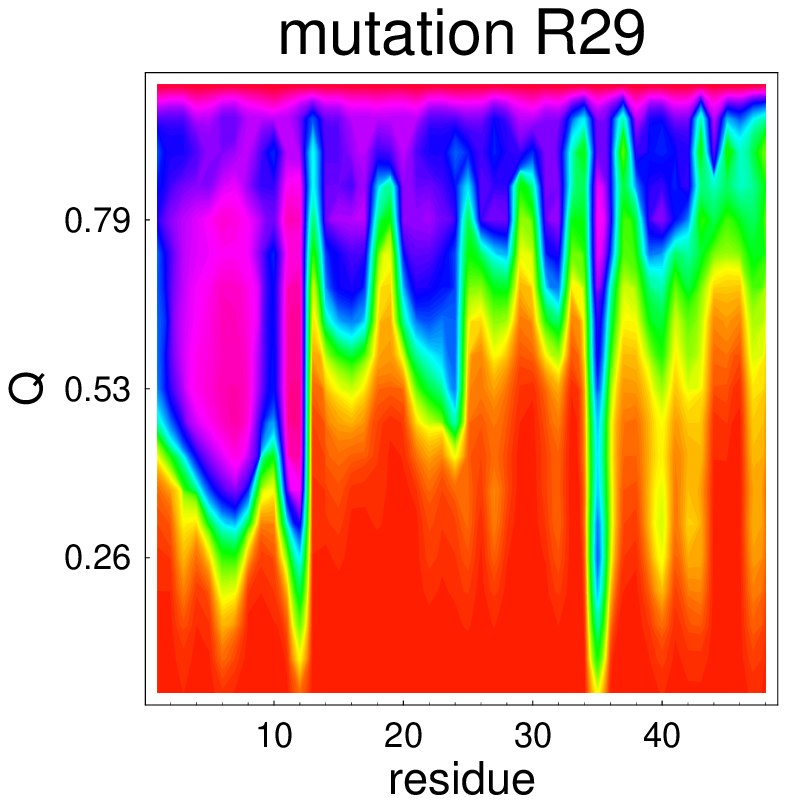}}}\hspace{0.7cm}}
{\rotatebox{0}{\resizebox{5cm}{5.0cm}{\includegraphics{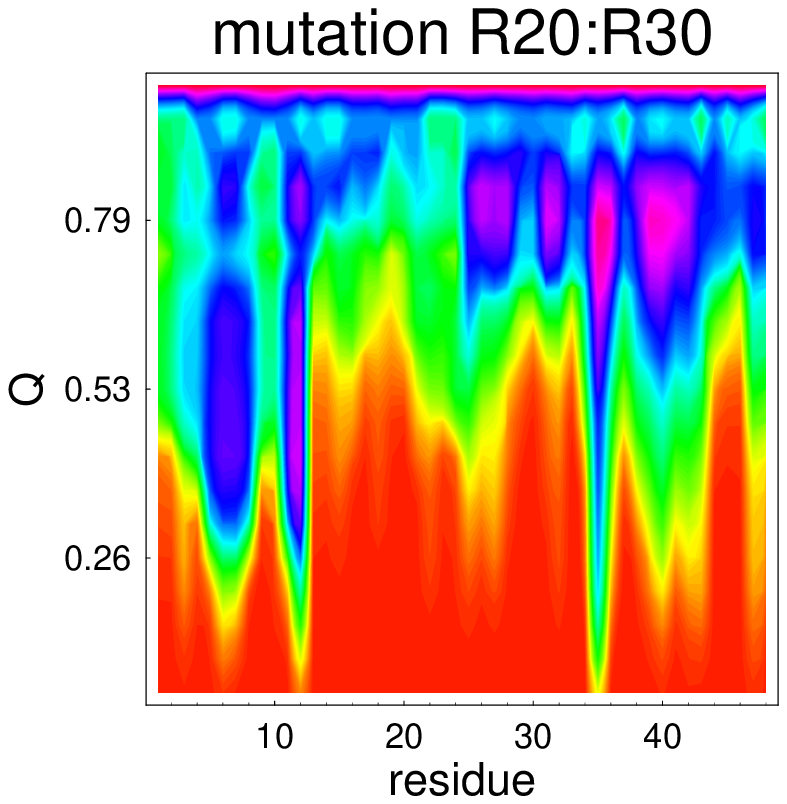}}}}\\
{\rotatebox{0}{\resizebox{7cm}{1.0cm}{\includegraphics {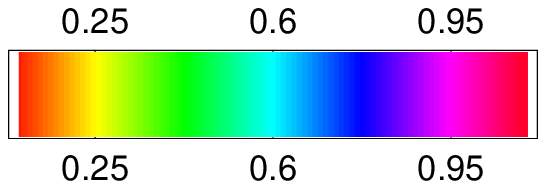}}}}
\caption{Folding patterns of geometry 1: probability that each residue is in its native environment along the reaction coordinate fraction of native contacts, $Q$, for the wild type (top left) and mutated sequences in geometry 1.}
\label{fig:no3}
\end{figure*}

\begin{figure*}
{\rotatebox{0}{\resizebox{5cm}{5.0cm}{\includegraphics
{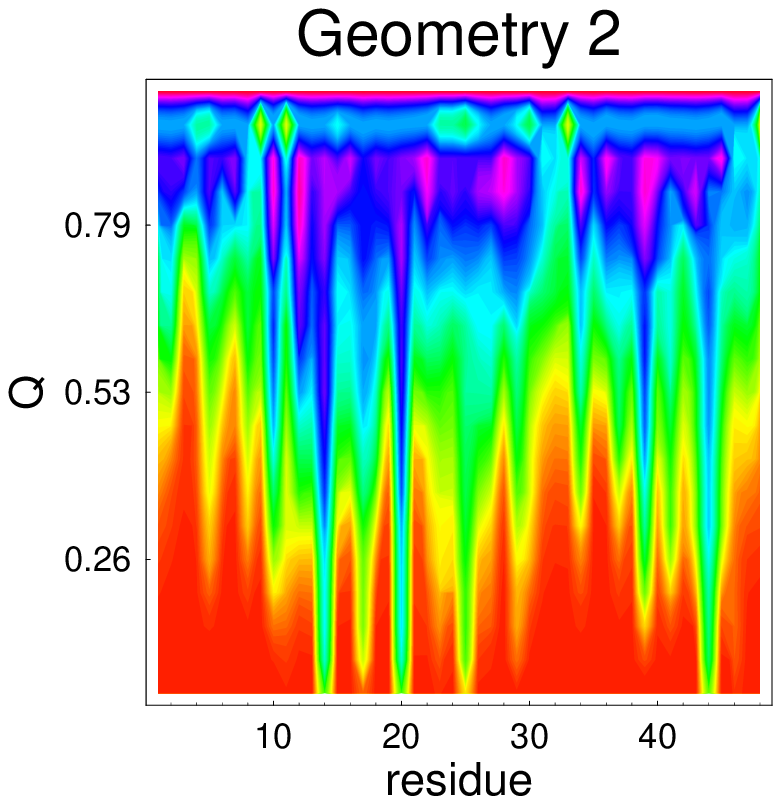}}}\hspace{0.7cm}} 
{\rotatebox{0}{\resizebox{5cm}{5.0cm}{\includegraphics
{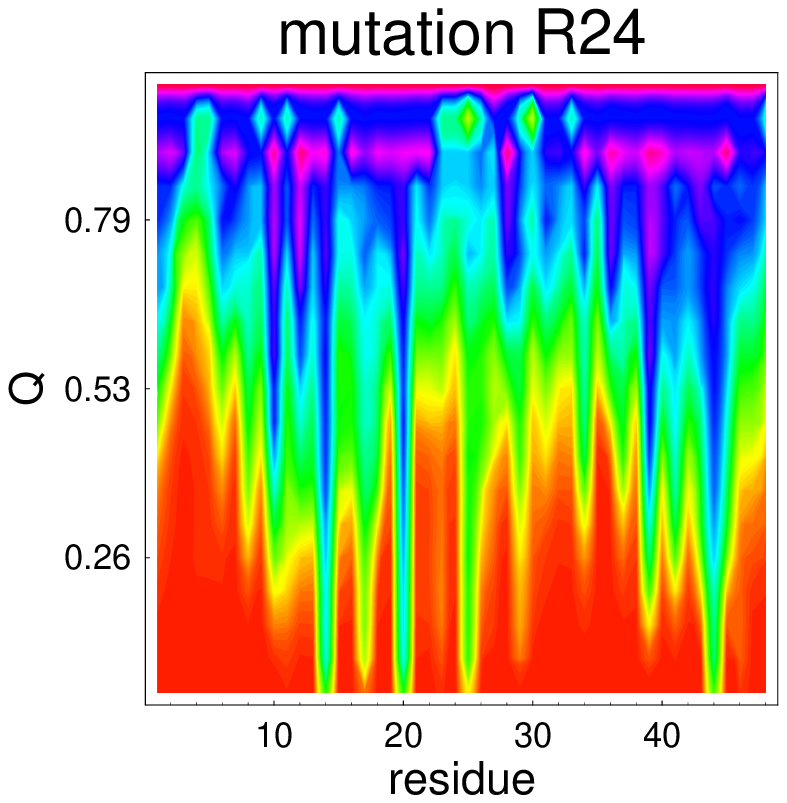}}}} \\
{\rotatebox{0}{\resizebox{5cm}{5.0cm}{\includegraphics
{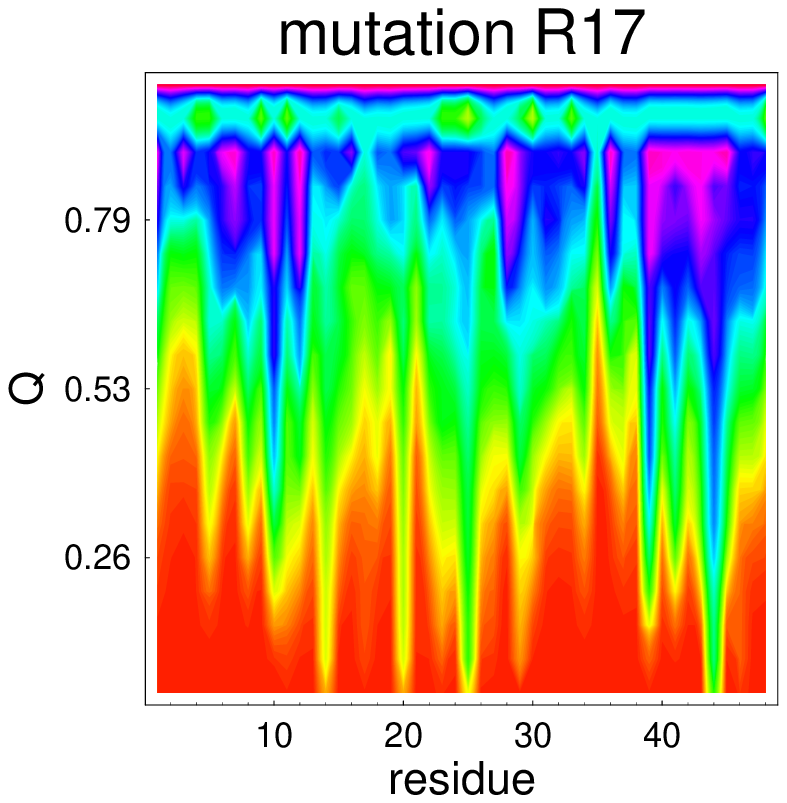}}}\hspace{0.7cm}} 
{\rotatebox{0}{\resizebox{5cm}{5.0cm}{\includegraphics
{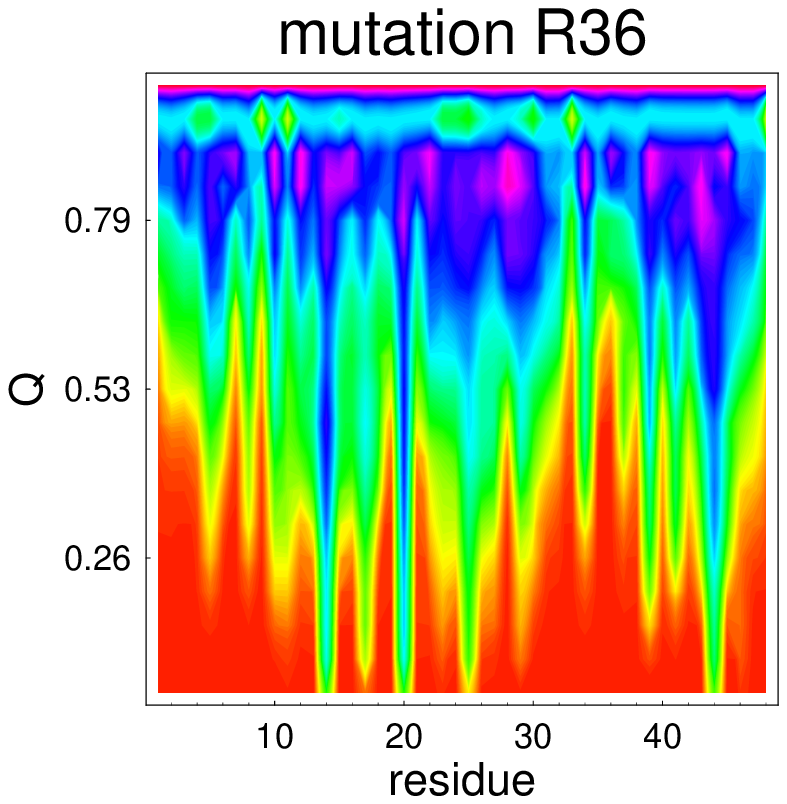}}}} \\
{\rotatebox{0}{\resizebox{5cm}{5.0cm}{\includegraphics
{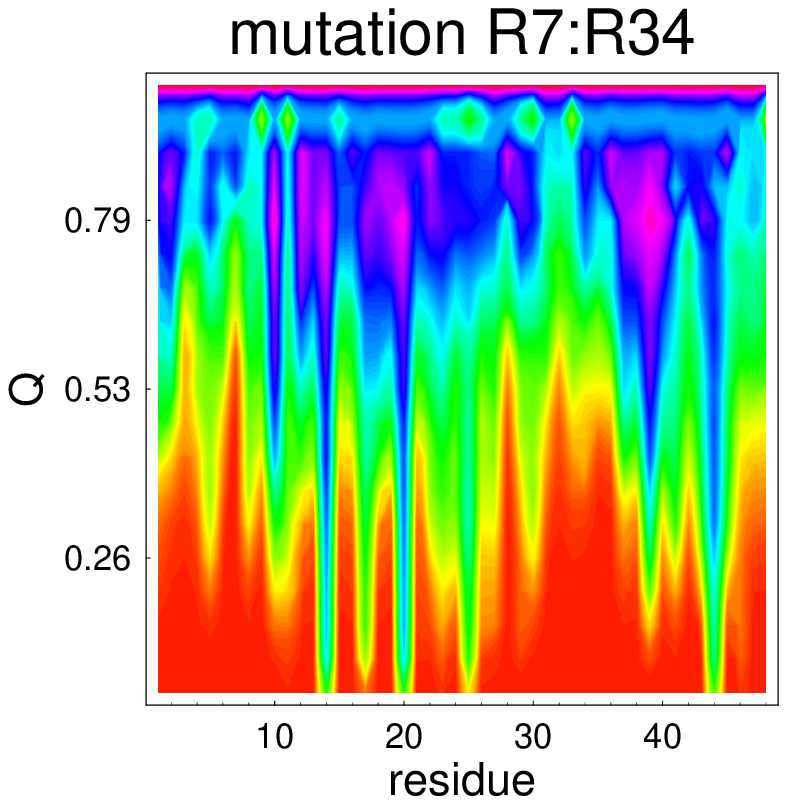}}}\hspace{0.7cm}} 
{\rotatebox{0}{\resizebox{5cm}{5.0cm}{\includegraphics {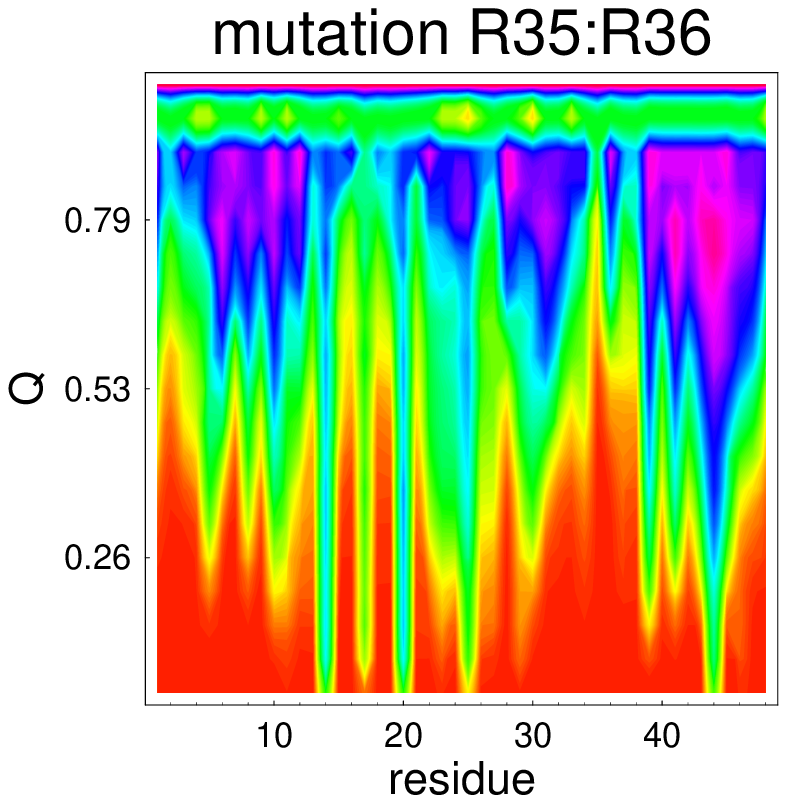}}}}\\
{\rotatebox{0}{\resizebox{7cm}{1.0cm}{\includegraphics{legend.eps}}}}
\caption{Folding patterns of geometry 2: probability that each residue is in its native environment along the reaction coordinate fraction of native contacts, $Q$, for the wild type (top left) and mutated sequences in geometry 2. }
\label{fig:no4}
\end{figure*}

 
Not surprisingly, the folding pattern of geometry 1 (Figure~\ref{fig:no3}, top left), i.e., the conformational changes leading to the native fold, is readily distinct from that exhibited by geometry 2 (Figure~\ref{fig:no4}, top left). For example, from early to mid folding (i.e., $0.26 <Q < 0.53$), there are more residues in geometry 2 with a higher probability of being in their native environment.  These probabilities decrease sharply immediately prior to collapse into the native fold (i.e., $0.79 < Q < 1.0$). On the other hand, for the vast majority of the residues belonging to geometry 1, the probability of being fully native increases in a rather continuous way from early to late folding.

\subsection{Native geometry and conformational plasticity}
\begin{figure}
{\rotatebox{0}{\resizebox{7cm}{7cm}{\includegraphics{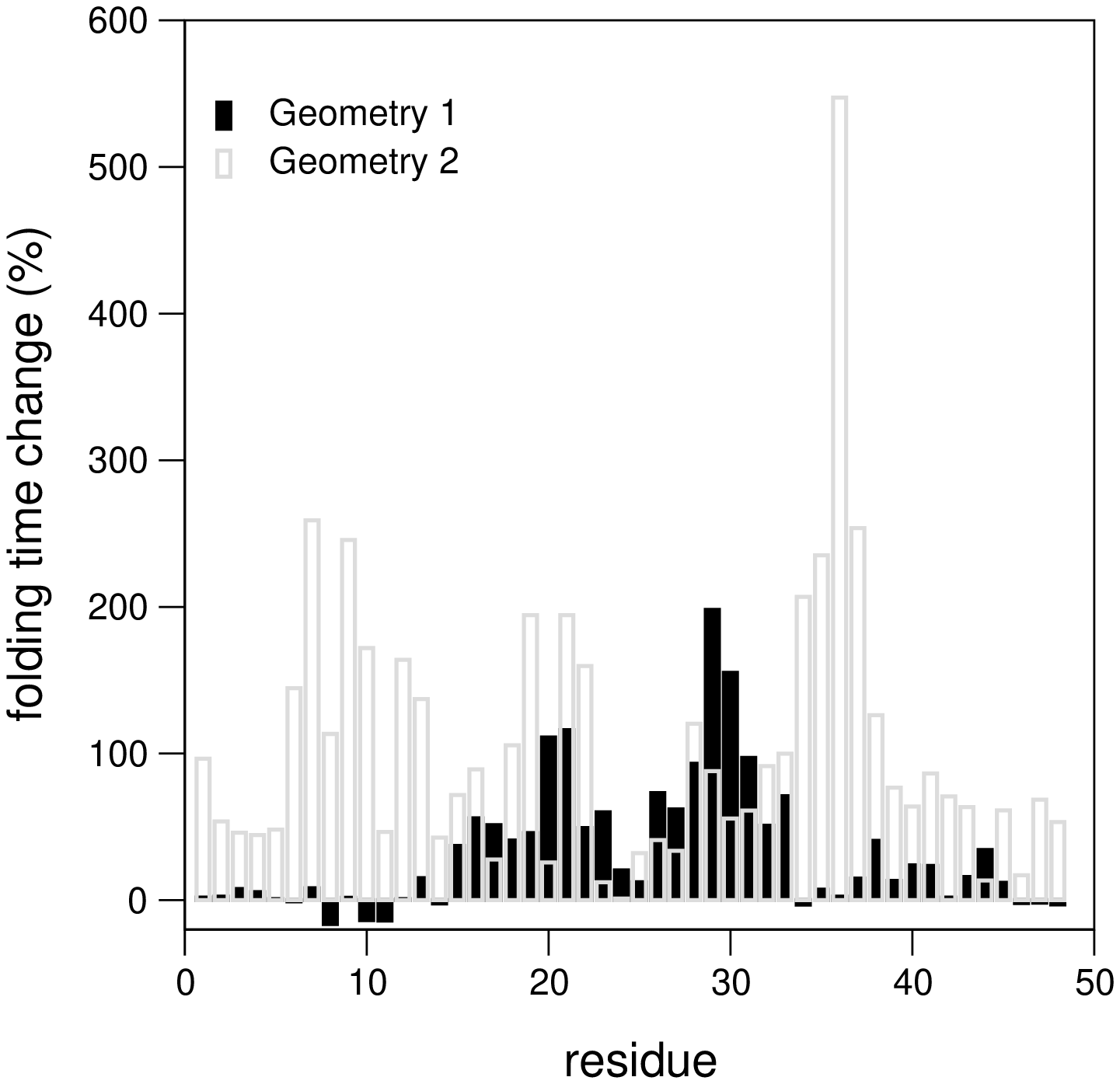}}}} 
\caption{Change in folding time relative to the wild-type protein resulting from single point mutations~\cite{FAISCA_2008}.}
\label{fig:no2}
\end{figure}

\begin{figure*}
{\rotatebox{0}{\resizebox{7cm}{7cm}{\includegraphics{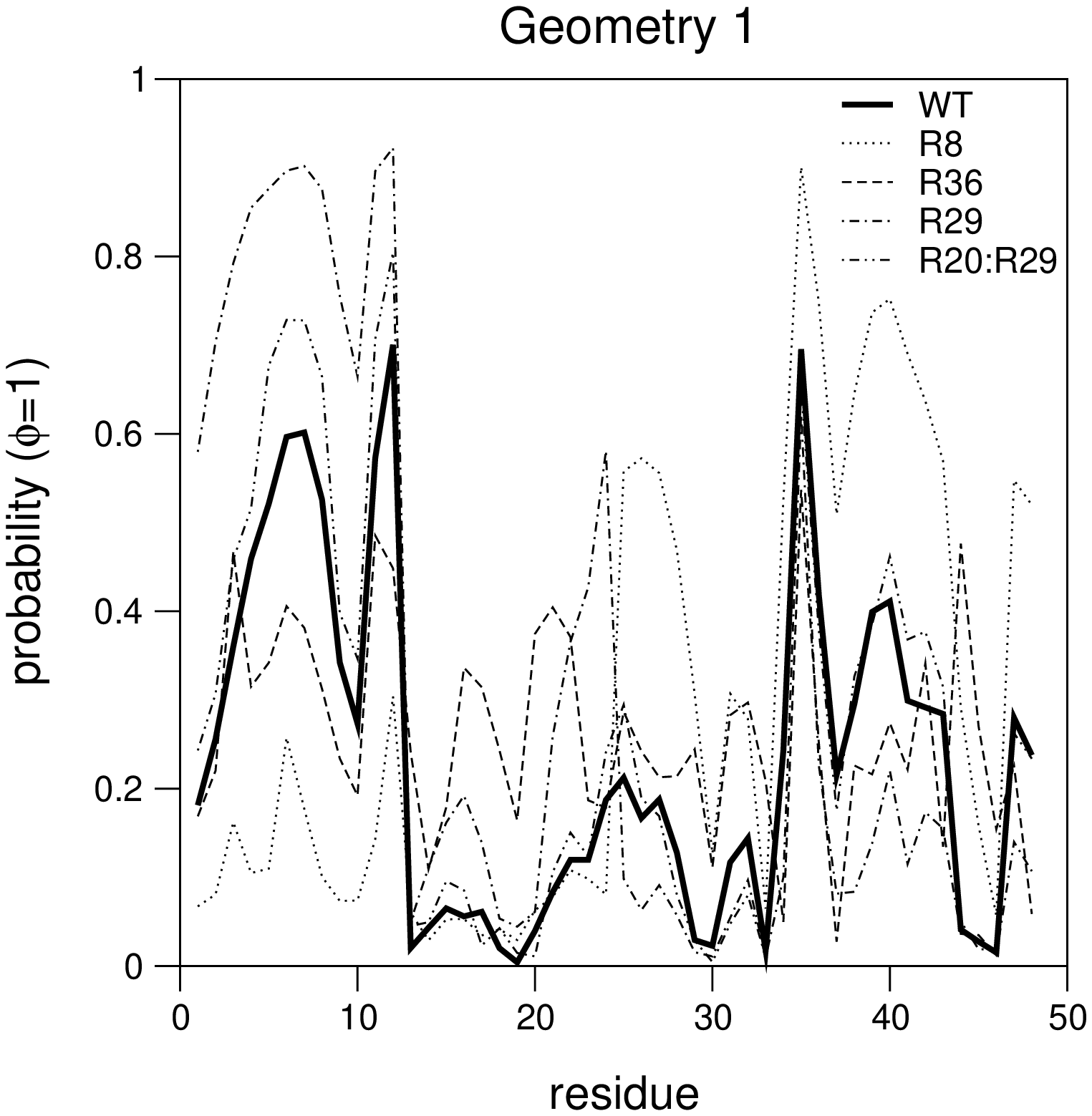}}}} 
{\rotatebox{0}{\resizebox{7cm}{7cm}{\includegraphics{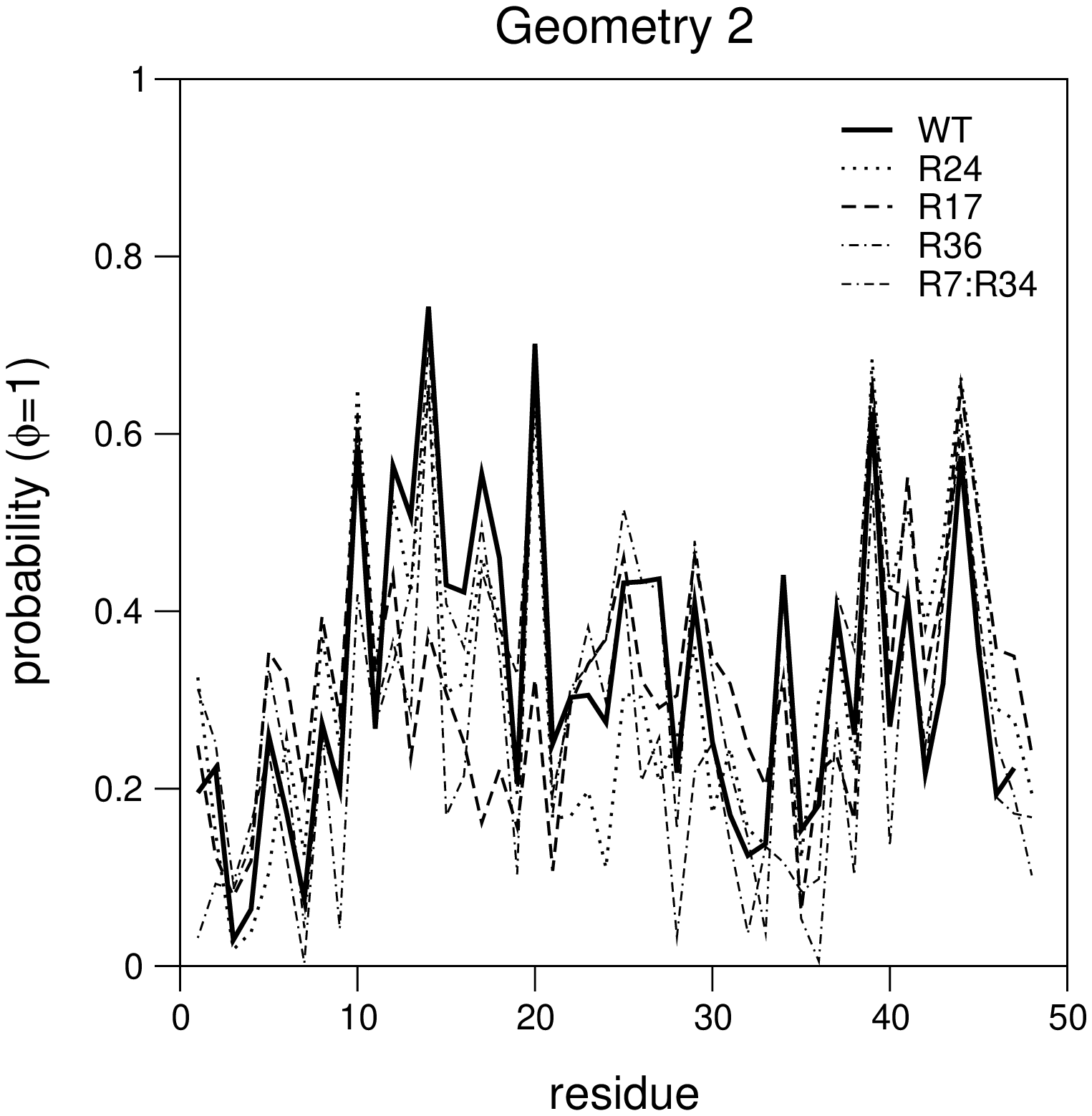}}}}
\caption{Probability that a residue is in its native environment in the wild-type transition state (TS) of geometry 1 (left) and geometry 2 (right). Also shown is the TS change induced by single and double-point  mutations. }
\label{fig:no5}
\end{figure*}

\begin{figure*}
{\rotatebox{0}{\resizebox{7cm}{7cm}{\includegraphics{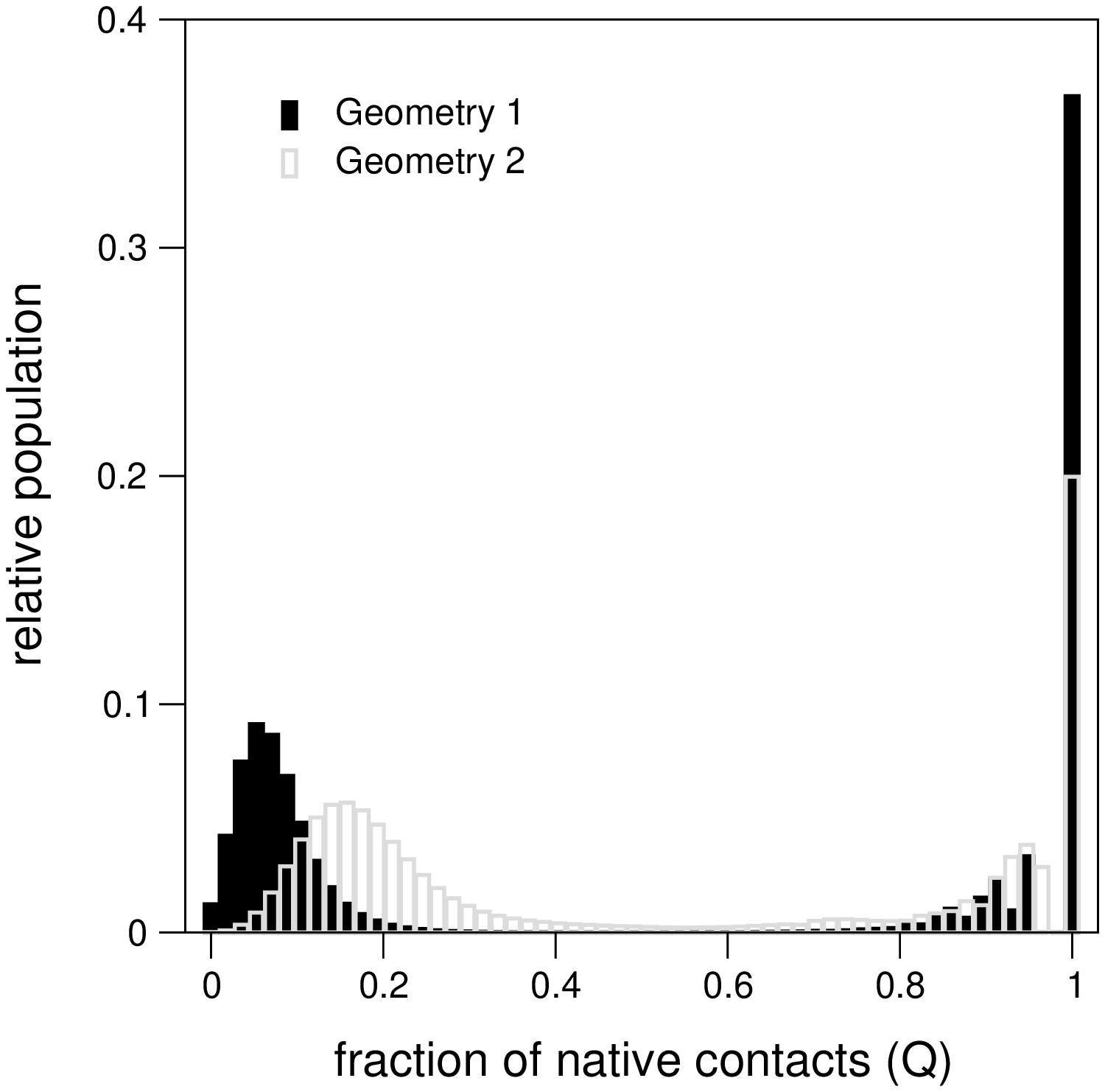}}}}
{\rotatebox{0}{\resizebox{7cm}{7cm}{\includegraphics{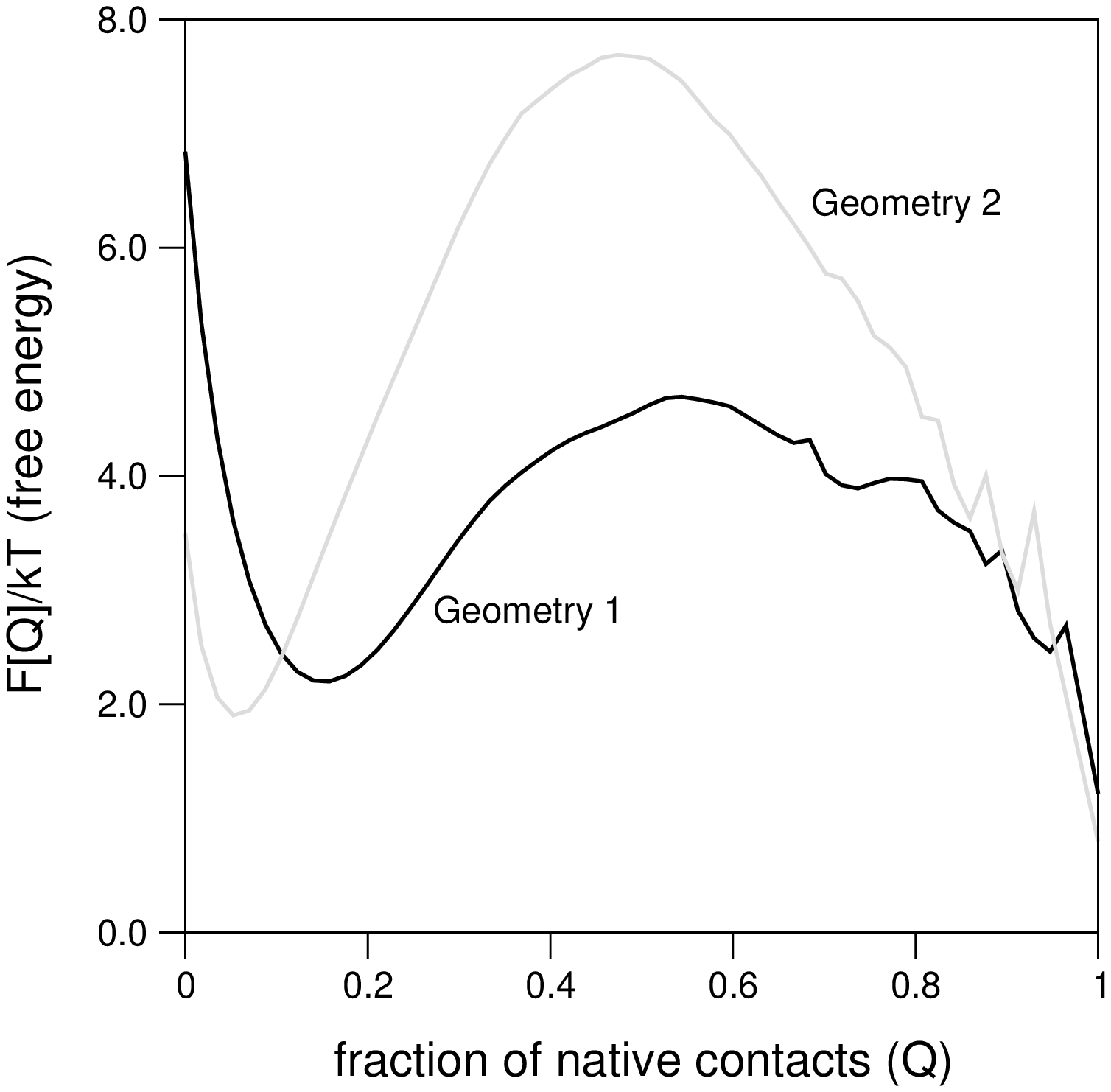}}}}
\caption{Cooperativity of the folding transition. Population histogram for the frequency of occurrence of conformations with fraction of native contacts $Q$, $P(Q)$ (left), and the free energy computed as $F(Q)=-kT \ln P(Q)$ (right). To compute $P(Q)$ data were averaged after collapse to the native state in long MC simulations lasting over $\sim$ $10^9$ MCS.}
\label{fig:no6}
\end{figure*}

Within the context of the G\={o} model a single-point mutation is equivalent to replace the set of native contacts established by the mutated residue with neutral contacts, i.e., contacts to which zero energy is ascribed. In a recent study we have mutated every single residue in each protein model and measured the change in the folding time of the mutant relative to the wild-type (WT) sequence (Figure~\ref{fig:no2}); an extensive number of double point mutations was also performed~\cite{FAISCA_2008}. 
We now ask how does the folding pattern of both protein models previously described alters upon 
mutation?  To answer this question we have selected single point mutations that produce different kinetic effects. For both geometries we have investigated the effect on conformational plasticity of the 
most deleterious and of the milder forms of single point mutations. We have also considered the impact on the conformational changes occurring during folding of the most deleterious double point mutations. Thus, for geometry 1 we have looked into the folding patterns obtained upon mutating residue  29 (which leads to the largest folding time increase of 200\%),  residues 5 and 36 (neutral mutations located below and above the sequence midpoint respectively, and also residue 8  (which decreases the WT protein folding time by 16.2\%) (Figure~\ref{fig:no2}). A simultaneous mutation on the residues 20 and 30, which is the most deleterious double point mutation for geometry 1, increases the folding time of the WT protein by almost two orders of magnitude~\cite{FAISCA_2008}. We have also investigated its impact on conformational plasticity. The pattern of the folding reaction associated with any one of these mutations is sharply different from that exhibited by the WT protein (Figure~\ref{fig:no3}), which is indicative that the less complex geometry has access to several folding pathways.  
For the more complex geometry 2, one has selected the mutation on residue 36 (which produces the largest increase in folding time of 600\%), on residue 24 (which is the only mutation for which a vanishingly small change of 1.2\% in the folding time is observed), and also on residue 17 (which leads to a mild increase of 27\% the WT protein's folding time). We have also investigated the effects of two double point mutations, namely on residues 7 and 34, and on residues 35 and 36. These mutations increase the WT protein's folding time by more than two orders of magnitude~\cite{FAISCA_2008}. A scenario, considerably different from that reported for geometry 1, is observed for geometry 2 (Figure~\ref{fig:no4}). Indeed, in this case, the folding pattern is considerably more robust, which suggests that the search for the native fold is constrained to follow a fixed sequence of conformational changes, i.e., that this geometry has a smaller conformational plasticity. \par

\subsection{Long-range contacts, the structure of the transition state and the cooperativity of the folding transition}

As discussed below, the large number of LR contacts in geometry 2 protects folding from perturbation,
which decreases the conformational plasticity, in two ways: i) by increasing the robustness of the transition state' structure against mutation and ii) by increasing the cooperativity of the folding transition. \par
Figure~\ref{fig:no5} (straight line) reports the probability that each protein residue is fully native in the TS ($Q\approx 0.5$) of geometry 1 (left) and geometry 2 (right), and how this probability changes upon mutation. The structure of the TS of geometry 2 is clearly more robust against mutation than that of geometry 1. Indeed, most mutations in geometry 1 lead to a large structural consolidation between residues 12 and 33. 
In geometry 2, the most deleterious double-point mutations disrupt networks of non-local LR contacts that form the core structure of the TS~\cite{FAISCA_2008}. Since the formation of LR contacts greatly restricts the number of conformations available to the folding chain, the establishment of LR contacts is generally entropically costly. Because the LR contacts are clearly dominant in the native fold of geometry 2 it is highly unlikely that upon the mutation of one or more of its residues, folding can occur through an energetically competitive TS, as this would imply changing the TS structure by increasing substantially its content in local (i.e., less entropically costly) native contacts~\cite{BAKER_nature, DOKHOLYAN}. Thus, for the non-local geometry 2, the robustness of its TS is a direct consequence of the fold's high content in LR contacts.\par 
In protein folding the formation (and breaking) of native contacts in a non-independent manner (i.e., cooperatively) results into the depletion of partially folded conformational states, which translates into a kinetic behaviour that fits remarkably well a two-state model. Indeed, at the transition midpoint of a two-state folding `reaction', half of the protein molecules in the test tube are folded and half of them are coil. Thus, microscopically, a cooperative folding transition is characterized by a bimodal distribution of protein molecules over energy, fraction of native contacts, $Q$, or any other observable parameter~\cite{PRIVALOV}. Data reported in Figure~\ref{fig:no6} (top) shows that at the transition temperature, $T_{m}$, the population distribution of the fraction of native contacts for geometry 2, $P(Q)$, is more strongly bimodal than that of geometry 1, showing that for geometry 2 the folding transition is clearly more cooperative~\cite{ABKEVICH1}. The stronger cooperative stabilization of the native state of geometry 2 relative to that of geometry 1 is more evident from the corresponding free energy curve, $F(Q)=-kT \ln P(Q)$, where a very high free energy maximum separates the native conformation from the of ensemble of unfolded conformers (Figure~\ref{fig:no6}, bottom)~\cite{GO}. Mechanistically, the strong cooperative folding of geometry 2 restricts the number of allowed conformational changes, and as a consequence, the number of alternative folding trajectories (i.e., the conformational plasticity) is smaller for this more complex geometry.

\section{Conclusion}

According to the statistical or landscape view of protein folding the native state can be reached from a myriad of microscopic parallel pathways. This `new' view contrasts the so-called traditional view that envisages folding as a Levinthal-like search, where the elements of the native structure assemble in a well-defined order. Thus, identifying the number of folding pathways accessible to a protein chain is a crucial question in molecular biology. Recent molecular dynamics simulations of the unfolding of CI2 at high temperature showed that the observation of sequential or multiple unfolding pathways depends on the `resolution', i.e. level of structural detail, at which one observes the unfolding process. Indeed, at the level of individual contact formation the unfolding of CI2 happens through highly parallel folding pathways while at the coarser level of contact clusters sequential folding events emerge~\cite{WEIKL_PROTEINS}. Experimentally, however, the observation of multiple folding pathways has proven a much more challenging task, and indirect evidence for alternative routes to the native state has been reported for engineered proteins only and the term folding or conformational plasticity was coined to denote the multiplicity of folding pathways identified for the perturbed (i.e., mutated) system. Folding plasticity for small perturbations (i.e., mild mutations) is taken as an indication that multiple folding routes exist for the wild-type protein.
Here we have investigated the effect of single- and double-point  mutations on the robustness of the folding reaction (and TS structure) leading to two distinct protein geometries. More specifically, 
we have investigated how the probability that each residue is fully native (i.e., that is has all its native contacts formed) evolves during the folding process, and how this evolution responds to mutation.   
Our findings suggest that folding to native geometries which are distinctively rich in long-range native contacts is more robust against mutation than folding to native geometries where the number of local native contacts is dominant. Indeed, `local' geometries seem to be able to find alternative folding routes in response to mutation while non-local geometries appear to behave in a Levintahl-like manner, their folding reaction being constrained to follow a fixed sequence of conformational changes, when monitored at the coarse level of contact cluster (i.e., the set of native contacts established by each residue). 
In other words, the conformational plasticity is expected to be small for target proteins where the number of non-local native contacts is sharply larger than the number of local contacts.
These findings are supported by experimental results reported for the $\lambda_{6-85}$-repressor (an $\alpha$-protein that is rich local contacts) and for protein CI2, and src SH3 domain ($\alpha$/$\beta$ proteins rich in non-local contacts); while the TSs of the latter were found to be largely invariant against single and double point mutations~\cite{FERSHT_JMB, BAKER2}, substantial changes were found for the former. Interestingly, the more complex native fold, i.e., the one that is rich is non-local native contacts is clearly less symmetric than the more local geometry, suggesting that a relation similar to that found by Klimov and Thirumalai between the symmetry of the tertiary structure and the folding plasticity~\cite{KLIMOV_JMB} may also hold for lattice proteins.\par
Mechanistically, the lower folding plasticity of the more complex native geometries is most possibly a direct consequence of their more cooperative folding transition, which results from a large content in long-range native contacts. In agreement with this explanation, experimental evidence for high conformational plasticity, resulting from a weak cooperative behaviour, was recently reported. Indeed, a change in the folding mechanism was found for two members of the {\it cks} family as a result of weakening the cooperativity of the core of the protein~\cite{SEELIGER}.\par
In our previous work~\cite{FAISCA_2008} we have concluded that native folds having a distinctively large number of non-local native contacts are more suitable targets for TS studies based on the use of $\phi$-value analysis. The results reported here strengthen this conclusion; indeed, due to their smaller folding plasticity, model proteins with a distinctively large number of LR contacts are preferable targets for TS studies based on the traditional interpretation of $\phi$-values. 
      
\clearpage

\section{\bf {Acknowledgments}}P.F.N.F. thanks Funda\c c\~ao para a Ci\^encia e Tecnologia (FCT) for financial support through grants SFRH/BPD/21492/2005 and POCI/QUI/58482/2004.


\clearpage
\begin{thebibliography}{99}


\bibitem{DOBSON_KARPLUS}{Dobson C. M., Karplus M. The fundamentals of protein folding: bringing together theory and experiment, Curr. Opinion Struct. Biol. (1999) 9:92-101.}
\bibitem{SHEA}{Weihua G., Shea J. E., Berry S. The physics of the interactions governing folding and association of proteins, Ann. N. Y. Acad. Sci. (2005) 1066:34-53.} 
\bibitem{JACKSON}{Jackson S. E. How do small single-domain proteins fold? Fold Des. (1998) 3:R81-91.}
\bibitem{PLAXCO} {Plaxco K.W., Simmons K.T., Ruczinski I., Baker D. Topology,
stability, sequence and length: Defining the determinants of two-state
protein folding kinetics, Biochemistry (2000) 39:11177-11183.}
\bibitem{GROMIHA}{Gromiha M.M. and Selvaraj S. Comparison between long-range interactions and contact order in determining the folding rate of two-state folders: Application of long-range order to folding rate prediction, J. Mol. Biol. (2001) 310:27-32.}
\bibitem{ZHOU}{Zhou H. and Zhou Y. Folding Rate Prediction Using Total Contact Distance, Biophys. J. (2002) 82:458-463.}
\bibitem{FERSHT_JMB}{Itzhaki L. S., Otzen D. E., Fersht A. R. The structure of the transition-state for folding of chymotrypsin inhibitor-2 analyzed by protein engineering methods: evidence for a nucleation-condensation mechanism for protein-folding, J. Mol. Biol. (1995) 254:260-288.}
\bibitem{FERSHT_BOOK}{Fersht A. Structure and Mechanism in Protein Science: A guide to enzyme catalysis and protein folding, W. H. Freeman (1998) 3rd Edition.}
\bibitem{WEIKL_PNAS}{Merlo C., Dill K.~A. and Weikl T.~R. $\phi$-values in protein-folding kinetics have energetic and structural components, Proc. Natl. Acad. Sci. USA 102:10171-10175 (2005).}
\bibitem{WEIKL_JMB}{Weikl T.~R. and Dill K.~A. Transition states in protein folding kinetics:
The structural interpretation of $\phi$-values, J. Mol. Biol. 365:1578-1586 (2007).}
\bibitem{WEIKL_BJ}{Weikl T.~R. Transition States in Protein Folding Kinetics: Modeling $\phi$-values of small $\beta$-sheet proteins, Biophysical Journal 94:929-937 (2008).}
\bibitem{FERSHT}{Fersht A.R. and Sato S. $\phi$-value analysis and the nature of protein-folding transition states, Proc. Natl. Acad. Sci. U.S.A. 101:7976-7981 (2004).}
\bibitem{MUNOZ}{Munoz V. Folding plasticity, Nature Struct. Biol. (2002) 9:792-794.}
\bibitem{OLIVEBERG_REV}{Lindberg M.O.,  Oliveberg M. Malleability of protein folding pathways: a simple reason for complex behaviour, Curr. Opin. Struct. Biol. (2007) 17:21-29.}
\bibitem{SERRANO}{Viguera A. R., Serrano L., and Wilmanns M. Different folding transition states may result in the same native structure, Nature Struct. Biol. (1996) 3:874-880.}
\bibitem{BAKER_1}{Grantcharova V.P., Baker D. Circularization Changes the Folding Transition State of the src SH3 Domain, J. Mol. Biol. 306 (2001) 555-563.}
\bibitem{OLIVEBERG}{Lindberg M., Tangrot J. and Oliveberg M. Complete change of the protein folding transition state upon circular permutation, Nature Struct. Biol. (2002) 9:818-822.}
\bibitem{HUBNER}{Hubner I.A., Lindberg M., Haglund E., Oliveberg M. and Shakhnovich E.I. Common motifs and topological effects in the protein folding transition state, J. Mol. Biol. (2006) 359:1075-1085.}
\bibitem{HUBNER_1}{Hubner I.A., Oliveberg M. and Shakhnovich E.I., Simulation, experiment, and evolution: Understanding nucleation in protein S6 folding, Proc. Natl. Acad. Sci. U.S.A. (2006) 101:8354-8359.}
\bibitem{CLEMENTI}{Matysiak S., Clementi C. Optimal combination of theory and experiment for the characterization of the protein folding landscape of S6: How far can a minimalist model go?, J. Mol. Biol. (2004) 343:235-248.}
\bibitem{PRIVALOV}{Makhatadze, G.I. and Privalov, P.L. Energetics of protein structure. Adv.
Protein Chem. 47:307-309 (1995).}
\bibitem{WEIKL_PROTEINS}{Reich L and Weikl T.R. Substructural Cooperativity and Parallel Versus Sequential Events During Protein Unfolding, Proteins 63:1052-1058 (2006).}
\bibitem{FERSHT1998}{Otzen D.E. and Fersht A.R. Folding of circular and permuted chymotripsin inhibitor 2: Retention of the folding nucleus, Biochemistry (1998) 37:8139-8146.}
\bibitem{BAKER2}{Grantcharova V.P., Riddle D.S. and Baker D. Long-range order in the src SH3 folding transition state, Proc. Natl. Acad. Sci. U.S.A. (2000) 97:7084-7089.}
\bibitem{OAS}{Burton R.E., Huang G. S., Daugherty M.A., Calderone T.L. and 
Oas T.G., The energy landscape of a fast-folding protein mapped by Ala-Gly substitutions
Nat. Struct. Biol. (1997) 4:305-310.}
\bibitem{KLIMOV_JMB}{Klimov D.K.,  Thirumalai D. Symmetric connectivity of secondary structure elements enhances the diversity of folding pathways. J. Mol. Biol. (2005) 353:1171-1186.}
\bibitem{GO}{Go N., Taketomi H. Respective roles of short- and long-range interactions in protein folding. Proc Natl. Acad. Sci. U.S.A. (1978) 75:559-563.}
\bibitem{BINDER}{D. P. Landau and K. Binder, A guide to Monte Carlo Simulations in statistical physics (Cambridge University Press, 2000)}
\bibitem{METROPOLIS}{N. Metropolis, A. W. Rosenbluth, M. N. Rosenbluth, A. H. Teller and E. Teller, Equation of state calculations by fast computing machines, J. Chem. Phys. (1953) 21:1087-1092.}
\bibitem{OLIVEBERG2}{Oliveberg M., Tan Y., Fersht A. R., Negative activation enthalpies in the kinetics of protein folding, Proc. Natl. Acad. Sci. U.S.A. (1995) 92:8926-8929.}
\bibitem{JCPSHAKH}{A. Gutin, A. Sali, V. Abkevich, M. Karplus, E.I. Shakhnovich,
Temperature dependence of the folding rate in a simple protein model:
Search for a ``glass'' transition, J. Chem. Phys. (1998) 108:6466-6483.}
\bibitem{CIEPLAK}{Cieplak M., Hoang T. X., Li M. S. Scaling of folding properties in simple models of proteins. Phys. Rev. Lett. (1999) 83:1684-1687.}
\bibitem{PFN1}{Faisca P.F.N., Ball R.C. Thermodynamic control and dynamical regimes in protein folding. J. Chem. Phys. (2002) 116:7231-7238.}
\bibitem{ABKEVICH1}{Abkevich V.I., Gutin A.M. and Shakhnovich E.I., Impact of local and non-local interactions on thermodynamics and kinetics of protein folding, J. Mol. Biol. 252 (1995) 460-471.}
\bibitem{VENDRUSCOLO}{Vendruscolo M., Paci E., Dobson C.M, and Karplus M., Three key residues form a critical contact network in a protein folding transition state, Nature (2001) 409:641-645.}
\bibitem{HUBNER_JMB}{Hubner I. A., Shimada J. and Shakhnovich E. I. Commitment and nucleation in the protein G transition state, J. Mol. Biol. 336:745-761 (2004).}
\bibitem{PACI_07}{Allen L.R. and Paci E., Transition states for protein folding using molecular dynamics and experimental restraints, J. Phys.: Condens. Matter 19 (2007) 285211}
\bibitem{DU}{Du R., Pande V.S., Grosberg A.Y., Tanaka T. and Shakhnovich E.S., On the transition coordinate for protein folding, J. Chem. Phys.(1998) 108:334-350.}
\bibitem{GILLESPIE}{Gillespie B. and Plaxco K.W., Non-glassy kinetics in the folding of a simple, single domain protein, Proc. Natl. Acad. Sci. U.S.A. (2000) 97:12014-12019.}
\bibitem{CHO}{Cho S., Levy Y. and Wolynes P.G., P versus Q, Structure reaction coordinates capture protein folding on smooth landscapes, Proc. Natl. Acad. U.S.A. (2006) 103:586-591.}
\bibitem{CLEMENTI1}{Clementi C., Jennings P.A. and Onuchic J.N. Prediction of Folding Mechanism for Circular-permuted Proteins, J. Mol. Biol. 311:879-890 (2001).}
\bibitem{FAISCA_2008}{Faisca P.F.N., Travasso R., Ball R.C. and Shakhnovich E.I. Identifying critical residues in protein folding: Insights from $\phi$-values and $P_{fold}$ analysis, J. Chem. Phys., {\it in press} (arXiv:0806.3064v1).}
\bibitem{SEELIGER}{Seeliger M.A., Breward S.E. and Itzhaki L.S. Weak cooperativity in the core causes a switch in folding mechanism between two proteins of the cks family, J. Mol. Biol. (2003) 325:189-199.}
\bibitem{BAKER_nature}{Baker D. A surprising simplicity to protein folding,  Nature (2000) 405:39-42}
\bibitem{DOKHOLYAN}{Dokholyan N.V., Buldyrev S.V., Stanley H.E. and Shakhnovich E.I., Identifying the protein folding nucleus using molecular dynamics, J. Mol. Biol. (2000) 296:1183-1188}



\end{thebibliography}
\end{document}